\documentclass[prb,amsmath,amssymb,notitlepage,reprint]{revtex4-1}

\newcommand{\dd}[1]{\mathrm{d}#1\,}

\renewcommand{\Re}{\mathop{\mathrm{Re}}}
\renewcommand{\Im}{\mathop{\mathrm{Im}}}

\DeclareMathOperator{\Tr}{Tr}
\DeclareMathOperator{\tr}{tr}

\DeclareMathOperator{\arsinh}{arsinh}

\newcommand{\bra}[1]{\langle{#1}\rvert}
\newcommand{\ket}[1]{\lvert{#1}\rangle}
\renewcommand{\vec}[1]{\bm{#1}}

\usepackage{graphicx}
\usepackage{hyperref}
\usepackage{xcolor}
\usepackage{units}
\usepackage{bm}

\begin{document}

\title{Superconducting size effect in thin films under electric field:
  mean-field self-consistent model}

\author{P.~Virtanen}
\affiliation{NEST, Istituto Nanoscienze-CNR and Scuola Normale Superiore, I-56127 Pisa, Italy}
\email{pauli.virtanen@nano.cnr.it}

\author{A.~Braggio}
\affiliation{NEST, Istituto Nanoscienze-CNR and Scuola Normale Superiore, I-56127 Pisa, Italy}

\author{F.~Giazotto}
\affiliation{NEST, Istituto Nanoscienze-CNR and Scuola Normale Superiore, I-56127 Pisa, Italy}

\begin{abstract}
  We consider effects of an externally applied electrostatic field on
  superconductivity, self-consistently within a BCS mean field model,
  for a clean 3D metal thin film. The electrostatic change in
  superconducting condensation energy scales as $\mu^{-1}$ close to subband edges as
  a function of the Fermi energy $\mu$, and follows 3D scaling
  $\mu^{-2}$ away from them. We discuss nonlinearities beyond gate effect, and
  contrast results to recent experiments.
\end{abstract}

\maketitle

\section{Introduction}

Quantum oscillations in superconducting properties due to size
quantization in thin films were predicted early
\cite{blatt1963-srs,falk1963-spb,shanenko2007-ost,shanenko2015-afs},
and they were later observed in metallic films
\cite{orr1984-tto,guo2004-smq,zhang2005-bso,bao2005-qse}.
Modification of superconducting properties by changing the electron
density by electrostatic fields was also observed,
\cite{mannhart1992-cis,mannhart1993-efe,ahn2003-efe,matthey2007-efe,ahn2006-emn,shvarts2005-mif}
and is best studied in high-Tc superconductors where the charge
density can be low enough to enable efficient gating.  Generally,
modifications of critical temperature $T_c$ and critical current $I_c$
have been reported.  Modification of $I_c$ only was also recently
reported in Refs.~\onlinecite{shvarts2005-mif} and
\onlinecite{simoni2018-msf,*paolucci2018-ues,*paolucci2018-mef} in metallic thin-film
samples, but the proper interpretation in the latter is still unclear.

Electrostatics of superconductors is an old problem (see e.g.
Ref.~\onlinecite{lipavsky2002-eps} for a historical review), and the
effect of electric fields on superconducting surfaces are
theoretically discussed in several works.
\cite{shapiro1984,shapiro1985,lee1996-iee,burlachkov1993-icc,lipavsky2006-glt,morawetz2008-sem,morawetz2009-dco,ummarino2017-pet}
In these, effects on the amplitude of superconductivity ($T_c$) are
usually related to modulation of electronic density of states (DOS),
which is also what contributes to the quantum size effects. A common
approach is to consider ``surface doping'' and assume the DOS is
modified within a Thomas--Fermi screening length from the
surface. Self-consistently screened calculations in superconductors
have been previously discussed in Refs.~\onlinecite{koyama2001-pev,koyama2003-foc,woods2018-eta}, in
a different context.  For the normal state, there is a large
literature on microscopic calculations with surface screening, which
are routine today e.g. using density functional
theory. \cite{lang1970-tm2,ummarino2017-pet} Modification of $I_c$ on
the other hand is often assumed to come from changes in the vortex
surface pinning potential.  \cite{mannhart1991-ief,shvarts2005-mif}

In a simple picture, a static electric field appears as a perturbation
of the potential that confines electrons within the thin film.  Static
fields generally extend up to a screening length from the surface, and
so their effect decreases towards high charge density. Although the
effects increase with the applied electric field, achievable field
magnitude is limited by electric breakdown (e.g. via field emission
\cite{gadzuk1973-fee}).

Electrostatic gating of superconductivity in the BCS mean field
picture relies on electron-hole asymmetry within an energy window
determined by the order parameter and Debye frequency centered at the
Fermi level.
\cite{adkins1968-bps,khomskii1992-crp,khomskii1995-cvh,otterlo1995-vdh}
In a simple clean thin-film model, strong asymmetry naturally exists
in the form of the steplike multiband 2D DOS, which also gives rise to
the quantum size effect, and the picture also extends to weakly
disordered samples. The only question is to what degree the DOS
asymmetry is retained, even though sharp features in the DOS are
smeared by disorder, \cite{belitz1994-amt} and when samples cannot be
significantly gated (metallic regime
\onlinecite{simoni2018-msf,*paolucci2018-ues,*paolucci2018-mef}),
since the Fermi level is not necessarily fixed at a sensitive point.
Regardless, sharp DOS features can increase the charge density range
in which electrostatic effects are large enough to be observed.
Motivated by the recent experimental results
\onlinecite{simoni2018-msf,*paolucci2018-ues,*paolucci2018-mef} where
large effects were seen, we revisit the problem.

In this work, we write down and solve a simple mean-field model for
superconductivity in thin films under electric fields, including
self-consistent screening. We point out connections between the
dependence of electrostatic energy on superconductivity and modulation
of superconductivity by fields, and discuss applicability of ``surface
doping'' models in this picture. We also discuss to what degree
nonlinear effects beyond linear electrostatic gating could appear in strong
fields.  We conclude that effects such as observed in
Ref.~\onlinecite{simoni2018-msf,*paolucci2018-ues,*paolucci2018-mef} likely are not
present in the model considered.

The manuscript is structured as follows. In Sec.~\ref{sec:mf} we introduce the
mean-field model considered and discuss results obtained
for the electric fields and modulation of superconducting properties.
Sec.~\ref{sec:conclusions} concludes with discussion.

\section{Mean-field model}

\label{sec:mf}

Self-consistent electrostatic screening and the size effect on
superconductivity in a clean superconducting metal in a simple
mean-field approximation is convenient to consider starting from a
Hartree--Bogoliubov free energy. It can be obtained
\cite{rice1967-sot,ambegaokar1982-qdo,otterlo1995-vdh,otterlo1999-dea}
by decoupling a long-ranged Coulomb and a (retarded) superconducting
contact interaction via Hubbard--Stratonovich transformations, and
considering only the classical saddle point in the static limit:
\begin{align}
  \label{eq:F}
  F[\Delta,\phi]
  &=
  -T \Tr \ln \mathcal{G}^{-1}
  \\\notag
  &
  +
  \int\dd{^3r}\Bigl(
  \rho \phi
  -
  \frac{\epsilon_0}{2}(\nabla\phi)^2
  +
  \int_0^{\frac{1}{T}}\dd{\tau}
  \frac{|\Delta(\tau)|^2}{\lambda(\tau)}
  \Bigl)
  \,,
  \\
  \label{eq:Ginv}
  \mathcal{G}^{-1}
  &=
  -i\omega + [\frac{\hat{k}^2}{2m} - U - \mu - e\phi]\tau_3 + \Delta(\omega)\tau_1
  \,.
\end{align}
Here, $\mathcal{G}$ is the electron equilibrium Green function, $U$ is
a background potential, $\mu$ chemical potential, $\phi$ is equivalent
to the static electric potential, $\Delta$ the superconducting order
parameter, and $\rho$ ion and external charge density.  The electron
charge is $-e$ and we use units with $\hbar=k_B=1$.
The first term in the free energy is the electronic contribution,
and the second part contains the electrostatic and superconducting
mean-field contributions.
In the absence of
currents and magnetic field, at saddle point with suitable gauge
$\Delta$ can be chosen real-valued and the values of vector
potential and phase are zero.  Above, $\phi$ has to be taken as the
saddle-point solution, which as typical for variational Poisson
does not minimize $F$.

Variations vs. $\phi$ and $\Delta$ give the Poisson and BCS
self-consistency equations:
\begin{gather}
  \label{eq:poisson}
  \begin{split}
  -\epsilon_0\nabla^2\phi(\vec{r})
  &=
  \rho(\vec{r}) - en_e(\vec{r})
  \\&
  =
  \rho(\vec{r}) + e T \sum_{\omega_n}\tr\tau_3\mathcal{G}(\vec{r},\vec{r},\omega_n)
  \,,
  \end{split}
  \\
  \Delta(\vec{r})
  =
  -
  \frac{1}{2}T\sum_{|\omega_n|<\omega_c}\lambda(\vec{r})\tr\tau_1\mathcal{G}(\vec{r},\vec{r},\omega_n)
  \,,
\end{gather}
where $\mathcal{G}$ satisfies the Gor'kov equations
$\mathcal{G}^{-1}\mathcal{G}=1$ under the self-consistent potentials.
We also here consider a BCS weak-coupling model, with
$\Delta(\omega)=\Delta\theta(\omega_c-|\omega|)$, with the coupling
$\lambda$ taken as constant and the cutoff $\omega_c$ similar to the
Debye frequency.  In bulk,
the BCS gap equation is then directly
$\Delta=2\omega_ce^{-1/(N_0{}\lambda)}$ with $N_0$ the DOS per spin at
Fermi level.

For uniform system, expanding $\mathcal{G}$ in
Eq.~\eqref{eq:poisson} to lowest order in $\phi$ results to
$\epsilon^{\rm RPA}(\vec{q}) q^2\phi(\vec{q}) = \delta\rho(\vec{q})$,
where $\epsilon^{\rm RPA}(\vec{q}) = \epsilon_0 -
\frac{e^2}{q^2}\Pi(\vec{q};\Delta)$ is the self-consistent static
dielectric function of a clean
superconductor. \cite{prange1963-dcs,seiden1966-cds} In this model,
the static fields are screened, and external charge density affects
the electronic DOS, but not the Coulomb effect
\cite{morel1962-css} to $\lambda$. The latter is due to considering mean-field with the
decoupling assumed; corrections appear from fluctuations of $\phi$
(see e.g. Ref.~\onlinecite{fischer2018-sdb} for explicit calculations),
or on mean-field level with different decoupling \cite{schulz1990-eas}.

\begin{figure}
  \includegraphics{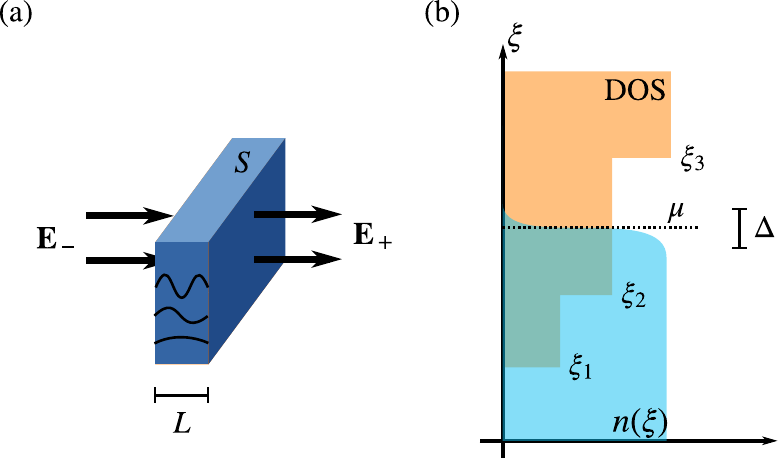}
  \caption{
    \label{fig:setup}
    (a)
    Schematic of superconducting quantum well of thickness $L$
    with infinite size in other directions,
    supporting several populated subbands,
    with electric fields $\mathbf{E}_\pm$ imposed on the surfaces.
    (b)
    Charge density~\eqref{eq:thinfilm-ne} in a superconductor is
    determined by the density of states and a occupation factor
    broadened by the superconducting interaction.
  }
\end{figure}

Aiming to describe effects on a qualitative level, we now consider a
simplified model, similar to those used in several previous studies of
the quantum size effect in superconducting thin
films. \cite{blatt1963-srs,shanenko2007-ost} A confining potential
$U(\vec{r})$ is taken to be an infinite quantum well at $|x|<L/2$,
which supports some number of populated 2D electronic subbands (see
Fig.~\ref{fig:setup}).  In a static problem without currents, the
electric field is perpendicular to the metal surface, and the problem
is inhomogeneous only in $x$-direction.  Moreover, we take as a
variational Ansatz $\Delta$ spatially constant \cite{falk1963-spb}
inside the well; the resulting energies will then be upper bounds to
the exact solutions.

With these assumptions, the problem is elementary and mostly given by
known results, \cite{anderson1959-tds,falk1963-spb}
and can be solved without further approximations.
First,
\begin{align}
  \mathcal{G}
  &=
  -
  \int_{-\infty}^\infty\dd{\xi}
  A_N(\vec{r},\vec{r}',\xi)
  \begin{pmatrix}
    \frac{u^2}{i\omega - \epsilon} + \frac{v^2}{i\omega + \epsilon}
    &
    \frac{uv}{i\omega-\epsilon} - \frac{uv}{i\omega+\epsilon}
    \\
    \frac{uv}{i\omega-\epsilon} - \frac{uv}{i\omega+\epsilon}
    &
    \frac{v^2}{i\omega - \epsilon} + \frac{u^2}{i\omega + \epsilon}
  \end{pmatrix}
  \,,
\end{align}
where $A_N$ is the normal-state spectral function (per spin),
$u,v=[\frac{1}{2}(1\pm\frac{\xi}{\epsilon})]^{1/2}$ and
$\epsilon=\sqrt{\xi^2+\Delta^2}$.  Due to the spatial symmetry, the
problem reduces to one dimension.  The normal-state DOS per volume is
$\nu(\xi)=\frac{2}{\mathcal{V}}\int\dd{^3r}A_N(\vec{r},\vec{r},\xi)=\sum_{n=1}^\infty\frac{m}{\pi
  L}\theta(\xi-\xi_n)$ where $\xi_n$ are the 2D subband edges and
$\mathcal{V}$ the film volume. The subbands and the potential $\phi$
are obtained from the Schr\"odinger--Poisson problem,
Eq.~\eqref{eq:poisson} with
\begin{gather}
  \label{eq:thinfilm-ne}
  n_e[\phi]
  =
  \sum_{n=1}^\infty 2m|\psi_n|^2\gamma(\xi_n)
  \,,
  \\
  [-\frac{1}{2m}\partial_x^2 - \mu - e\phi(x)]\psi_n = \xi_n\psi_n
  \,,
  \quad
  \psi_n(\pm\frac{L}{2})=0
  \,,
\end{gather}
where $\psi_n(x)$
are the transverse wave functions of the 2D subbands.
Here, $\gamma$ describes the contribution to the charge from each
subband:
\begin{align}
  \label{eq:nxi}
  n(\xi)
  &=
  f_0(\xi)
  +
  T\sum_{|\omega|<\omega_c}\frac{\xi\Delta^2}{(\omega^2+\xi^2+\Delta^2)(\omega^2+\xi^2)}
  \,,
  \\
  \label{eq:gamma}
  \gamma(\xi)
  &=
  \int_{\xi}^\infty\frac{\dd{\xi'}}{2\pi}n(\xi')
  \overset{T=0,\omega_c=\infty}{\simeq}
  \frac{\sqrt{\xi^2 + \Delta^2} - \xi}{4\pi}
  \,,
\end{align}
where $n(\xi) \to u^2 f_0(\epsilon) + v^2 (1 - f_0(\epsilon))$ for
$\omega_c\to\infty$ and $f_0$ is the Fermi function.
\cite{tinkham1996-its} The occupation factor $n$ is broadened by the
interactions in a window $\Delta$ around the Fermi level, with the
deviation from the Fermi function starting to decay more rapidly
beyond the interaction range at $|\xi|\gtrsim\omega_c$.  Variations in
the DOS within this window contribute to the charge
response of the amplitude of superconductivity (see
Fig.~\ref{fig:setup}).
\cite{adkins1968-bps,hong1975,khomskii1992-crp}

To be specific, we assume an external charge density outside the
sample (e.g. capacitor plates with constant charge density) such that
the amplitudes of the electric fields at the surfaces are fixed,
$-\partial_x\phi(\pm\frac{L}{2})=E_\pm$.  Numerically, the nonlinear
Poisson problem can be solved iteratively, \cite{eyert1996-csm} for
a fixed value of $\Delta$.

The condensation energy
$f_{ns}(\Delta)=(F[\Delta,\phi_*[\Delta]]-F[0,\phi_*[0]])/\mathcal{V}$
per volume for fixed $\Delta$ now depends only on the density of
states. \cite{anderson1959-tds,falk1963-spb}
Via direct calculation,
\begin{align}
  f_{ns}(\Delta)
  &=
  \frac{1}{\mathcal{V}}
  \int_0^\Delta
  \dd{\Delta}
  \frac{d}{d\Delta}
  F[\Delta,\phi_*]
  \,,
\end{align}
where we note that
$\frac{d}{d\Delta}F[\Delta,\phi_*]=\partial_\Delta{}F[\Delta,\phi_*]$
at the saddle point $\phi_*$. Further, \cite{falk1963-spb}
\begin{align}
  f_{ns}(\Delta)
  &=
  \frac{\Delta^2}{\lambda}
  -
  \int_0^\Delta\dd{\Delta}
  \sum_{|\omega|<\omega_c}
  \int_{-\infty}^\infty\dd{\xi}
  \frac{\nu(\xi)T\Delta}{\omega^2+\xi^2+\Delta^2}
  \\
  &\equiv
  \frac{\Delta^2}{\lambda}
  -
  \frac{m}{2\pi L}
  \int_0^\Delta\dd{\Delta}\Delta
  \sum_{n=1}^\infty
  g(\frac{\xi_n}{\Delta})
  \,,
  \\
  g(y)
  &=
  \frac{T}{\Delta}
  \sum_{|\omega|<\omega_c}
  \int_{y}^{\infty}\dd{x}
  \frac{1}{x^2 + 1 + (\omega/\Delta)^2}
  \,.
\end{align}
Here,
$g(y)\to
\int_y^\infty\dd{x}\frac{1}{\sqrt{1+x^2}}\frac{2}{\pi}\arctan\frac{\omega_c/\Delta}{\sqrt{1
    + x^2}}$ for $T=0$ and further $g(y)\to\arsinh(\omega_c/\Delta)-\arsinh{}(y)$ for
$\omega_c\gg\Delta$, $T=0$. 
For $T=0$ and $\omega_c\to\infty$,
\begin{align}
  f_{ns}(\Delta)
  &\to
  \frac{\Delta^2}{\lambda}
  -
  \frac{m\Delta^2}{4\pi L}
  \sum_{\xi_n<\omega_c}
  [
    \eta(\omega_c/\Delta) - \eta(\xi_n/\Delta)
  ]
  \,,
\end{align}
where $\eta(y)=\arsinh y + (\sqrt{y^2+1}-|y|)y$.
The self-consistent value $\Delta_*$ is attained at
a solution of $f_{ns}'(\Delta_*)=0$.

Separating out an electrostatic contribution by subtracting the result
for some reference potential $\phi_0$:
\begin{align}
  \delta{}f_{ns}
  &\equiv
  f_{ns}(\Delta;\phi_*) - f_{ns}(\Delta;\phi_0)
  \\
  \label{eq:F-est}
  &=
  -\frac{m}{2\pi L}
  \int_0^\Delta\dd{\Delta}\Delta
  \sum_{n=1}^\infty
  \Bigl(
  g(\frac{\xi_n}{\Delta})
  -
  g(\frac{\xi_n^{(0)}}{\Delta})
  \Bigr)
  \\
  \label{eq:delta-fns-expansion}
  &\simeq
  \frac{2m}{L}\sum_{n=1}^\infty\delta\gamma(\xi_n)\delta\xi_n
  \,,
  \qquad
  T=0,
  \;\omega_c\to\infty
  \,,
\end{align}
where
$\delta\gamma(\xi)\equiv\gamma(\xi,\Delta)-\gamma(\xi,\Delta=0)$
from Eq.~\eqref{eq:gamma}, and
$\delta\xi_n\equiv\xi_n-\xi_n^{(0)}$.  The result~\eqref{eq:F-est}
includes both gating \cite{blatt1963-srs,shapiro1984,shapiro1985} and
any nonlinear effects (e.g. energy associated with quantum
capacitance) in strong electric fields.  Note that the above
electrostatic energy contribution depends on the electric fields only
via $\xi_n=\xi_n[\phi]$, an exact statement in the model here.

It is also possible to express the electrostatic energy directly in
terms of the self-consistent electric field, at small field
strengths. Consider expansion of the electronic energy around a
reference electric potential, considering small $\phi_1=\phi-\phi_0$
and $\rho_1=\rho-\rho_0$:
\begin{equation}
  \label{eq:G-expansion}
  \begin{split}
  -T\Tr\ln\mathcal{G}^{-1} + T\Tr\ln\mathcal{G}^{-1}\rvert_{\phi=\phi_0}
  \\
  =
  \int\dd{^3r}(-e)n_e[\Delta,\phi_0](\vec{r})\phi_1(\vec{r})
  \\
  +
  \frac{1}{2}
  \int\dd{^3r}\dd{^3r'}e^2\Pi[\Delta,\phi_0](\vec{r},\vec{r}')\phi_1(\vec{r})\phi_1(\vec{r}')
  +
  \ldots
  \,,
  \end{split}
\end{equation}
where $n_e$ is the electron density and $\Pi$ the density response
function.  Solving the resulting saddle-point equation for $\phi_1$
and substituting the solution into $F$ gives, after integration by parts:
\begin{align}
  \label{eq:poisson-energy}
  f[\Delta]
  &=
  f[\Delta,\phi_0]
  +
  \frac{1}{\mathcal{V}}
  \int\dd{^3r}
  \Bigl(
  \rho_{1} \phi_0
  +
  \frac{1}{2}
  \rho_{1}
  \phi_{1,*}
  \Bigr)
  \\
  \label{eq:poisson-energy-2}
  &=
  f[\Delta,\phi_0]
  +
  \sum_\pm
  \frac{\mp{}\epsilon_0E_\pm}{L}[\phi_0 + \frac{1}{2}\phi_{1,*}]_{x=\pm\frac{L}{2}}
  +
  C
  \,,
  \\
  \label{eq:poisson-pt}
  \rho_{1}
  &=
  -\epsilon_0\nabla^2\phi_{1,*}
  -
  \int\dd{^3r'}e^2\Pi[\Delta,\phi_0](\vec{r},\vec{r}')\phi_{1,*}(\vec{r}')
  \,,
\end{align}
where $C$ is independent of $\Delta$. In this order of expansion in
small $\phi_1$, the additional electrostatic field energy
in~\eqref{eq:poisson-energy} coincides with the standard
expression. The linear term $\sim\rho_1\phi_0$ describes a gate effect
on superconductivity, which in this approach we see to be related to
the $\Delta$-dependence of the equilibrium potential $\phi_0$.  Using
Eq.~\eqref{eq:poisson-pt} the quadratic part can be expressed as
$\sim\phi_1\epsilon^{\rm RPA}\phi_1$.  It corresponds to a (quantum)
capacitance modulation \cite{morawetz2008-sem,morawetz2009-dco} by
superconductivity.  The result~\eqref{eq:poisson-energy} can be
directly used for computing $\delta{}f_{ns}(\Delta)$ (if
$\delta\phi\equiv{}\phi[\Delta]-\phi[0]$ is known) and is equivalent
with~\eqref{eq:F-est} in the small-field limit.  However, due the
$\Delta$-dependence of $\phi_0$ it is not necessarily very practical
to compute, as solving the nonlinear Poisson problem is still required.
However, the above expressions can be used as a consistency check.

As noted above, we consider charge density $\rho=\rho_1+\rho_0$ where
$\rho_1$ outside the sample fixes the electric field at the surface.
Finally, we need to specify the background (``ion'') charge density
$\rho_0$.  The electric potential due to $\rho_0$ together with $U$ gives
the pseudopotential for the electron system.
\footnote{ The confining potential $U$ is not necessary for the
  formulation, except for enabling the use of the constant-$\Delta$
  approximation, see Appendix~\ref{app:conf}.  } For simplicity,
unless otherwise mentioned, below we assume
$\rho_0=en_e[\Delta=0,\phi=0,\mu]$, which results to $\phi_0=0$ as the
solution in the normal state, and $\mu$ becoming the parameter that
fixes the charge density in the normal state.  This is of course a
crude toy model of the surface electron behavior, even within
Hartree-type models \cite{lang1970-tm2}, but likely modifies mainly the precise
positions of the subbands but not the main
qualitative features of the effect of the screening of external
charges on superconductivity.

\subsection{Size effect in electric field}

\label{sec:qsize}

In the same way as the variation in thickness,
\cite{blatt1963-srs,falk1963-spb} gating by a surface electric field
can in principle make a single subband edge $\xi_n$ to cross the Fermi
level, which results to a jump in superconducting properties.  Such
response can be larger than in bulk material, and is not captured by
``surface doping'' models often used for the electric field effect,
\cite{shapiro1984,shapiro1985} where the LDOS $\nu(x,\xi)$ is assumed
to be modified in a surface layer of thickness of a screening length
$\lambda_{TF}$ according to bulk relations.  In addition, the field
screening is not exactly Thomas-Fermi type, but this causes less
relevant changes than the difference in the DOS.

The order of magnitude of $\delta{}f_{ns}$ can be estimated in a
Thomas--Fermi approximation. Taking
$\phi(L/2 + x')\simeq{} -E\lambda_{TF}e^{x'/\lambda_{TF}}$ for $x'<0$,
$\lambda_{TF}=\sqrt{\epsilon_0/(e^2\nu_F)}$,
\begin{align}
  \label{eq:dxi-p1}
  \delta\xi_n \simeq
  \bra{n}(-e)\phi\ket{n}
  =
  \frac{\lambda_{TF}^2}{L}eE_+ q(2\lambda_{TF}k_n)
  \,,
\end{align}
where $k_n=\pi{}n/L$ and $q(z)=z^2/(1+z^2)$. From Eq.~\eqref{eq:delta-fns-expansion}, and
keeping only the smallest $|\xi_n|<\omega_c$,
\begin{align}
  \label{eq:dfns-approx}
  \delta f_{ns}
  \simeq
  |f_{ns,3D}|
  \frac{
    2eEa_0
  }{
    \sqrt{\xi_n^2+\Delta^2}+|\xi_n|
  }
  \frac{
    \pi^2
  }{
    4(k_FL)^2
  }
  q(2\lambda_{TF}k_n)
  \,,
\end{align}
where $f_{ns,3D}=-\frac{1}{4}\frac{mk_{F}}{\pi^2}\Delta^2$ is the bulk
3D condensation energy and $a_0=4\pi\epsilon_0/(me^2)$ the Bohr
radius.  The above result is valid in the leading order in $\Delta$,
as $\phi$ is assumed to be independent of it.  The factor
$q(2\lambda_{TF}k_n)$ in reality depends on details of the screening,
and below we consider it as a constant of order of magitude 1.

Including the next-order eigenvalue perturbation $\delta\xi_n^{(2)}$
in~\eqref{eq:F-est} and considering terms of order $E^2$
gives the second-order correction,
\begin{align}
  \label{eq:dfns2-approx}
  \delta f_{ns}^{(2)}
  &\simeq
  \frac{m}{L}
  \sum_{k,n=1}^\infty
  \frac{\delta\gamma(\xi_n)-\delta\gamma(\xi_k)}{\xi_n - \xi_k}
  |\bra{k}(-e)\phi\ket{n}|^2
  \,,
\end{align}
where $n=k$ means the limit $\xi_k\to\xi_n$. This energy contribution
is associated with the change $\Pi[\Delta,0]-\Pi[0,0]$
(c.f. Eq.~\eqref{eq:G-expansion},~\eqref{eq:Pi-expression}) in the
static Lindhard function \cite{prange1963-dcs}.  However, it is of the
same order in $\Delta$ as the change $\Pi[0,\phi_0[\Delta]]-\Pi[0,0]$
due to the $\Delta$-dependent shift in the self-consistent equilibrium
potential, which we have neglected here.  As a consequence,
Eq.~\eqref{eq:dfns2-approx} is not the only contribution to the $E^2$
term, and solving the self-consistent electrostatic problem is in
general required. \footnote{Alternatively, one
  could base the expansion~\eqref{eq:G-expansion} around the
  normal-state potential $\phi_0=0$ and include a third-order term
  $\int\Gamma_3(x,x',x'')\phi(x)\phi(x')\phi(x'')$.  } Conversely,
calculation of the effect of superconductivity on the dielectric
function requires taking the self-consistency of
$\Delta=\Delta_*[\phi]$ into account. \cite{seiden1966-cds}

\begin{figure}
  \includegraphics{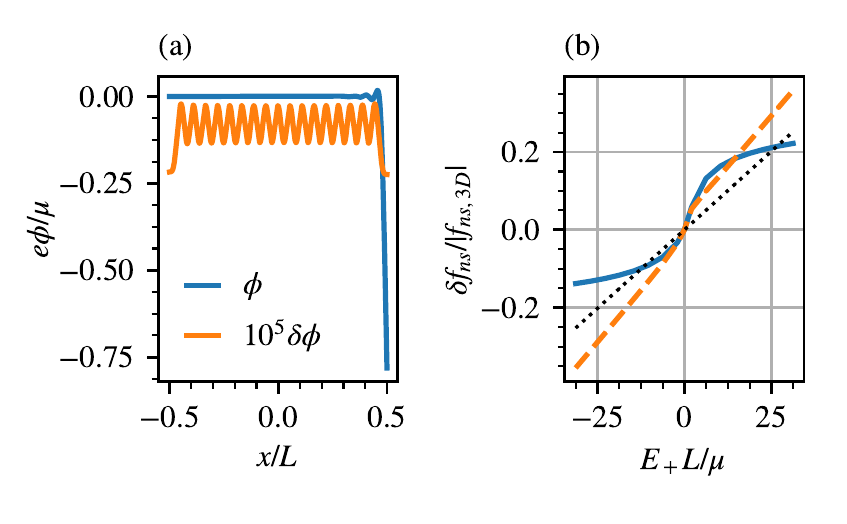}
  \caption{
    \label{fig:phi}
    (a)
    Self-consistent electric potential $\phi$ and its modulation
    $\delta\phi=\phi(\Delta)-\phi(\Delta=0)$ for $E_-=0$, $E_+=\unit[3.8]{V/nm}$,
    $L=\unit[10]{nm}$, $\mu=\xi_{18}^{(0)} + 0.5/(mL^2)=\unit[1.22]{eV}$,
    $\Delta=\unit[760]{\mu{}eV}$, $\omega_c=\unit[34]{meV}$, $T=0$.
    (b)
    Change $\delta{}f_{ns}$ in the condensation energy at fixed $\Delta$,
    in units of the 3D bulk condensation energy $f_{ns,3D}=-\frac{1}{4}\nu_{3D}(\mu)|\Delta|^2$.
    Results from Eq.~\eqref{eq:F-est} (solid), the small-field
    expression~\eqref{eq:poisson-energy-2} (dashed), and Eq.~\eqref{eq:dfns-approx}
    with $q(z)=1$ (dotted) are shown.
  }
\end{figure}

The overlap factor $q$ above depends on how accurate the Thomas-Fermi
screening assumption is close to the surface. For the simple problem
here, we can solve the Poisson equation numerically. Such a solution
is illustrated in Fig.~\ref{fig:phi}a for $\lambda_{TF}\ll{}L$.  Since
$\lambda_{TF}\sim{}k_F$, screening is not fully exponential, but the
electric potential exhibits $1/k_F$ oscillations.  The correction
$\delta\phi=\phi(\Delta)-\phi(\Delta=0)$ to the equilibrium
electrostatic potential from superconductivity is small in the high
charge density regime considered here. The chemical potential is chosen to be
close to a subband edge in the figure.

The corresponding dependence of $\delta{}f_{ns}$ on the electric field
magnitude is shown in Fig.~\ref{fig:phi}b, together with the
corresponding result from Eq.~\eqref{eq:dfns-approx}.  The
electrostatic energy expression~\eqref{eq:poisson-energy-2} is also
shown, and coincides with the exact result in the small-field regime.
Generally, the electric field effect is appreciable only for
$E_+L\gtrsim{}\mu$. In the estimate from Eq.~\eqref{eq:dfns-approx},
we here set $q(z)=1$, to account for the expectation that likely for
the true screening length $\lambda_{TF}k_F\gtrsim1$.  The second-order
correction~\eqref{eq:dfns2-approx} is neglibile for these parameters,
being higher order in $\lambda_{TF}^2eE/(L\omega_c)$, and the
nonlinearity visible in the result originates from $g(\xi)$.

\begin{figure}
  \includegraphics{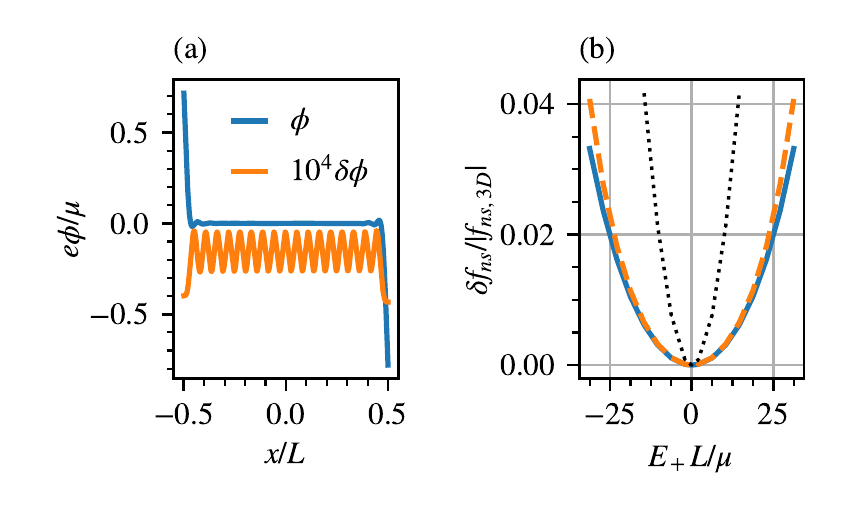}
  \caption{
    \label{fig:phi-sym}
    Same as Fig.~2, for the symmetric field configuration $E_+=E_-$.
    Dotted line indicates Eq.~\eqref{eq:dfns2-approx}.
  }
\end{figure}

The linear gating effect can be suppressed with a charge-neutral field
configuration $E_+=E_-=E$, which corresponds to an experiment using
floating gate electrodes (i.e. placing the system inside a plate
capacitor).  The result from Poisson equation for this case is shown
in Fig.~\ref{fig:phi-sym}, together with the result for
$\delta{}f_{ns}$. Imposing the field on both sides produces a larger
$\delta\phi$.  However, as the linear contribution to the free energy
cancels, the modulation $\delta{}f_{ns}$ arises from the next-order
effect and is an order of magnitude smaller than with the gate effect.
Although the energy can still be expressed also via
Eq.~\eqref{eq:poisson-energy-2}, the eigenvalue perturbation
result~\eqref{eq:dfns2-approx} does not agree as well, as expected.

\begin{figure}
  \includegraphics{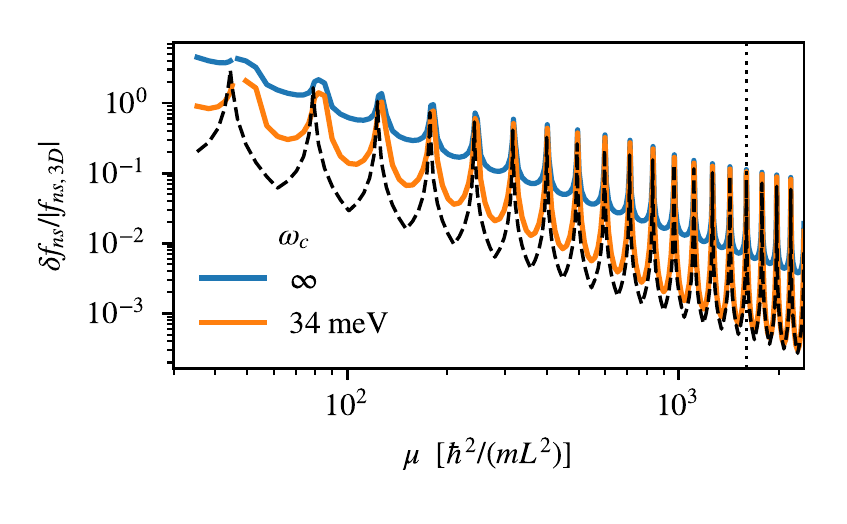}
  \caption{
    \label{fig:dfns-mu}
    Electrostatic condensation energy increase $\delta{}f_{ns}$ vs $\mu$
    for $E_+=\unit[0.76]{V/nm}$, other parameters as in Fig.~\ref{fig:phi}.
    Dashed line indicates Eq.~\eqref{eq:dfns-approx}, taking $q(z)=1$.
  }
\end{figure}

Whether electrostatic effects are significant depends on how large the
modulation $\delta{}f_{ns}$ is compared to $f_{ns}$.  The dependence
of their ratio on the chemical potential, and hence charge density, is
shown in Fig.~\ref{fig:dfns-mu}, at a relatively large external field.
The magnitude of $\delta{}f_{ns}$ depends strongly on whether the
chemical potential is located near a band bottom, where the effect is
amplified (see Fig.~\ref{fig:setup}), which produces the oscillations
visible in Fig.~\ref{fig:dfns-mu}.  When $\mu$ is close to a subband bottom, the
magnitude appears to be captured well by Eq.~\ref{eq:dfns-approx} (dashed line).
When the chemical potential is not close to a band bottom, depending
on the ratio between the subband spacing and the cutoff $\omega_c$,
the electric field effect can vary by order of
magnitude.
Note that as long as $|\xi_n|\ll\omega_c$ for some $n$, the result is
dominated by the smallest $\xi$ and the cutoff $\omega_c<\infty$ is of
limited importance.  The sum~\eqref{eq:F-est} is convergent also for
$\omega_c\to\infty$.  However, these results are based on the simple
weak-coupling model for superconductivity, and the precise shape of
the modulation may be sensitive to details of the
interaction. Regardless, from the above results one can see that the
relative magnitude at resonance scales as $\propto{}\Delta/\mu$, and
not as $(\Delta/\mu)^2$ as one would expect for the amplitude response
in 3D bulk \cite{prange1963-dcs,seiden1966-cds}. Away from the subband
edge resonances, $\delta{}f_{ns}\propto(\Delta/\mu)^2$.

\begin{figure}
  \includegraphics{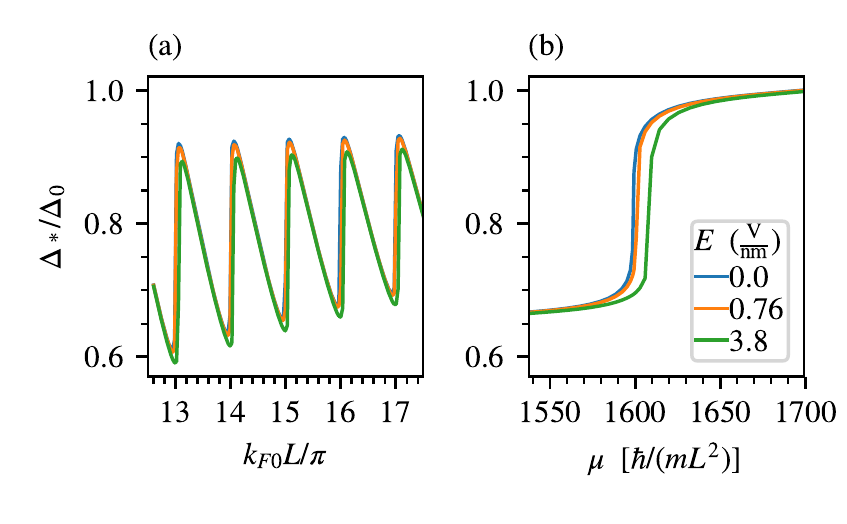}
  \caption{
    \label{fig:DeltaL}
    Self-consistent $\Delta_*(T=0)$ vs (a) $L$ and (b) $\mu$,
    with $g=N_{0,3D}\lambda=0.14$ fixed and other parameters as in Fig.~\ref{fig:phi}.
    Here, $\Delta_0=2\omega_ce^{-1/g}$.
  }
\end{figure}

The self-consistent value of $\Delta_*$, $f_{ns}'(\Delta_*)=0$, is
shown in Fig.~\ref{fig:DeltaL}a as a function of the film thickness,
showing the well-known quantum size
effect. \cite{blatt1963-srs,falk1963-spb} The corresponding dependence
on the chemical potential is shown in Fig.~\ref{fig:DeltaL}b, for
several values of the external electric field.  In this figure, it is
obvious that the electrostatic field simply gates the system: the size
effect physics is dominated by the $\xi_n$ closest to the chemical
potential, so that the gate-induced shift $\delta\xi_n$ is identical
to a chemical potential shift $-\delta\mu$.

The above discussion can be compared to a surface doping model, where the DOS is
assumed to change in a surface layer of thickness $\lambda_{TF}$, and
the system is considered a superconducting bilayer in the Cooper
limit.  In such a model,
$\delta{}F_{ns}/F_{ns}\sim{}-\frac{\lambda_{TF}}{L}\delta\frac{1}{\nu_F\lambda}\sim{}\frac{\delta\nu_{F}\lambda_{TF}}{\lambda
  \nu_F^2L}$.  The general form of Eq.~\eqref{eq:dfns-approx} can then
be recovered by including (ad-hoc) the main features of the multiband DOS in $\delta\nu_F$.
This can be done by writing
$\frac{\delta\nu_F}{\nu_F}=\frac{\pi}{k_FL}
\partial_{\xi_n}\theta_\Delta(\xi_n)\lambda_{TF}eE$, where $\theta_\Delta(\xi)$ is a
broadened unit step function with width $\Delta$.  For the problem
here, although the actual form of $\delta\nu(x,E)$ is different, this
simpler model captures the main effects.  Surface doping models have indeed
been successful in understanding previous experimental results.
\cite{mannhart1991-ief}

\subsection{Superfluid weight}

The effect of the electrostatic field on the phase fluctuations
can be studied via the superfluid weight $D_{ij}^s$,
\cite{scalapino1993-ims}
which describes the free energy cost of
superflow $\Delta(\vec{r})\propto{}e^{2i\vec{A}\cdot\vec{r}}$:
\begin{align}
  F[\Delta_*,\phi_*,A]
  =
  F[\Delta_*,\phi_*,0] + \frac{\hbar^2\mathcal{V}}{2}D_{ij}^sA_iA_j + \ldots
  \,.
\end{align}
The ``vector potential'' $\vec{A}$ describing the superflow
can be introduced in Eq.~\eqref{eq:Ginv} by
replacement $\hat{\vec{k}}\mapsto\hat{\vec{k}}+\vec{A}$.
The calculation of $D^s$ is standard for multiband BCS superconductor.
Since
the current operators in $y/z$ directions do not here couple different
bands, the result for $i,j\in\{y,z\}$ is
$D^s_{ij}=\delta_{ij}\sum_n{}n_s(\xi_n)/(mL)$ where $n_s(\xi_n)$ is the
BCS superfluid density: \cite{scalapino1993-ims}
\begin{align}
  n_s(\xi)
  &=
  2m
  \int_{\xi}^\infty\frac{\dd{\xi'}}{2\pi}[n(\xi') + (\xi'-\xi)f_0'(\epsilon')]
  \,,
\end{align}
where $n(\xi)$ is given by Eq.~\eqref{eq:nxi} and
$\epsilon'=\sqrt{(\xi')^2+\Delta^2}$. As well known,
$n_s(\xi)\to{}n_e(\xi)=2m\gamma(\xi)$ at $T=0$.  The electrostatic
modulation of the superfluid stiffness is then similar to that of the
charge density, i.e., small in the metallic regime. Similar conclusion
then applies to the phase stiffness, and quite likely
also to the phase-slip energy barrier \cite{langer1967-irt}. These
results however apply in the clean limit.

\section{Discussion and conclusions}

\label{sec:conclusions}

We discussed an elementary BCS/Hartree--Bogoliubov mean
field model for the size effect under self-consistent electrostatic
fields in superconducting thin films, and studied it based on
numerically exact solutions. As the size modulation in superconducting
properties decays relatively slowly with increasing charge density, it
increases the response to applied electric fields, effectively
changing the small parameter from $(\Delta/\mu)^2$ to $\Delta/\mu$ for
fine-tuned values of $\mu$, also in films thick compared to the
screening length.

The mean-field approach likely is not useful in describing atomically
thin, or strongly disordered and resistive samples, where fluctuation
effects matter.  Phase--plasmon fluctuation effects in principle can
be included in the approach above in a standard way by expanding in
$\Re\Delta$, $\Im\Delta$ and $V=-i\phi$ around the mean-field solution.  A
priori, in view of some existing results,
\cite{fischer2018-sdb,otterlo1999-dea} however, it's not clear why
such corrections would depend strongly on the external electric field.

Large electrostatic size effects in thin-film systems are expected to
be visible mainly in relatively low charge densities,
e.g. semiconducting materials.  As noted in previous works,
\cite{mannhart1991-ief} it appears likely this is a main effect in
high-Tc superconductors.  The modulation of screening by
superconductivity will also appear in proximity systems, e.g. in
semiconductor/superconductor hybrids \cite{vuik2016-eee,woods2018-eta}
recently considered as Majorana fermion platforms.

With regard to the large modification of superconducting critical
current by electric fields reported in
Refs.~\onlinecite{simoni2018-msf,*paolucci2018-ues,*paolucci2018-mef}, it
then appears somewhat less likely these results can be understood with
electrostatic effects of the type discussed above.  At metallic
densities $\Delta/\mu\sim{}10^{-4}$, electrostatic effects in the
model here, even at a sharp DOS feature, likely can only give
$|\delta{}f_{ns}/f_{ns,3D}|\lesssim10^{-2}$, which is too small to
cause large measurable effects.  It appears unlikely this is easily
rectified by lifting some of the approximations we made.  This is
simply a manifestation of the ``Anderson theorem'':
\cite{anderson1959-tds} the amplitude of conventional
superconductivity is insensitive to time-reversal symmetric
perturbations, and suppressing it requires perturbations large
compared to $\mu$, which are usually not achievable in the metallic
regime below the electrical breakdown field. Also, as the linear gate
effect generally should dominate nonlinearities, whether
superconductivity is suppressed or enhanced depends on the sign of the
electric field, quite unlike in
Refs.~\onlinecite{simoni2018-msf,*paolucci2018-ues,*paolucci2018-mef}.  Previously,
reduction in the critical current by an applied field was attributed
to modification of vortex pinning.
\cite{mannhart1991-ief,shvarts2005-mif} In
Ref.~\onlinecite{simoni2018-msf} effects appear also in aluminum
strips with lateral size $\lesssim\xi$, making this explanation less
favorable. In the clean-limit model here, it also appears unlikely the
phase slip rates would be significantly affected.

In summary, we considered effects of electrostatic fields on
superconductivity self-consistently within a BCS model, connected
them to questions about electrostatic energy, and commented on their
relation to recent experiments.  We obtain results for the size and
external electric field modulation of superconductivity, and contrast
results to a surface doping model.  Expanding about this mean field
solution, considering electric field effects on phase slips and phase
fluctuations is possible. Experimentally, the effects are best visible
in low charge density systems, e.g. semiconductor hybrid structures.

\bibliography{screenpaper}

\appendix

\section{Density response function in thin film}
\label{app:densresp}

The static density response in a superconducting infinite potential
well can be found, in a situation translationally invariant vs. $y$
and $z$ (i.e. response to a charge sheet).  First, we have
\begin{align}
  \label{eq:Pi-basic}
  \Pi(x,x')
  &=
  T
  \sum_{\omega_n}
  \tr \mathcal{G}(x,x')\tau_3\mathcal{G}(x',x)\tau_3
  \\
  &=
  2
  \int_{-\infty}^\infty\dd{\xi_1}\dd{\xi_2}A_N(\vec{r},\vec{r}';\xi_1)A_N(\vec{r},\vec{r}';\xi_2)^*
  \\\notag
  &\qquad\times
  \frac{n(\xi_1,\Delta) - n(\xi_2,\Delta)}{\xi_1 - \xi_2}
  \,,
\end{align}
where the trace and the Matsubara sum has been evaluated, and
$n(\xi)=u_\xi^2 f_0(\epsilon_\xi) + v_\xi^2 (1 - f_0(\epsilon_\xi))$.
The normal-state spectral function for a potential well is
\begin{align}
  A_N(x,x';\xi)
  =
  \sum_{p=1}^\infty
  \frac{2}{L}
  \sin[k_p(x+\frac{L}{2})]\sin[k_p(x'+\frac{L}{2})]
  \delta(\xi - E_p)
  \,,
\end{align}
where $k_p=\frac{\pi p}{L}$, $E_p=\frac{k_p^2}{2m}$. Then we have,
\begin{align}
  \label{eq:Pi-expression}
  \Pi(x,x')
  &=
  \frac{8m}{L^2}
  \sum_{pq=1}^\infty
  \sin[k_p(x+\frac{L}{2})]\sin[k_p(x'+\frac{L}{2})]
  \\\notag
  &\times
  \sin[k_q(x+\frac{L}{2})]\sin[k_q(x'+\frac{L}{2})]
  \\\notag
  &\times
  \frac{\gamma(E_p-\mu,\Delta) - \gamma(E_q-\mu,\Delta)}{E_p - E_q}
  \,,
\end{align}
which can be evaluated. Here, the terms $p=q$ imply the limit
$E_p\to{}E_q$.

\section{Confining potential}
\label{app:conf}

In a more realistic model than in the main text, we would set $U=0$
and the electrons would be confined in the metal film due to the
attractive potential from the ionic charge density $\rho_0>0$.
However, in such calculations the simplifying assumption of a
spatially constant $\Delta$ is not permissible, as discussed below.

The charge density in uniform 3D metal for $\mu\to-\infty$
(i.e. deep in the vacuum), with constant $\Delta$, is
\begin{align}
  \rho_e(\mu,T=0,\Delta)
  &=
  \frac{(2m)^{3/2}}{2\pi^2}\int_{-\mu}^\infty\dd{\xi}\sqrt{\mu+\xi}n(\xi)
  \\\notag
  &\simeq
  \frac{(2m)^{3/2}}{16\pi}\frac{\Delta^2}{\sqrt{-\mu}}
  \,.
\end{align}
The corresponding Poisson equation in a Thomas-Fermi approximation becomes
\begin{align}
  \partial_x^2\phi(x)
  &\simeq
  e\epsilon_0^{-1}\rho_e(\mu-\phi(x),T=0,\Delta)
  \simeq
  \frac{a}{\sqrt{\phi(x)}}
  \,,
  \\
  \Rightarrow
  \phi(x)
  &=
  \Bigl(\frac{3\sqrt{a}x}{2}\Bigr)^{4/3}
  \,,
  \quad
  \rho_e(x)\propto{}x^{-2/3}
  \,.
\end{align}
From the solution, we find the electrostatic field fails to confine
the ``superconducting'' electrons, and an infinite amount of total
charge $\int_{x_0}^{\infty}\dd{x}\rho_e(x)$ leaks to the vacuum, which
is unphysical.  In the exact solution, the mean field
$|\Delta(\vec{r})|$ would decrease simultaneously with the density,
providing a stronger electrostatic confinement.  Although the details
of this surface effect appear sensitive to the external electric
field, it appears unlikely it is important for the stability of
superconductivity in the bulk.

\end{document}